\documentclass[preprint]{ptephy_v2}  
\preprintnumber{XXXX-XXXX} 

\usepackage{lineno}
\usepackage{hyperref}

\usepackage{xcolor}
\usepackage{bm}
\usepackage[normalem]{ulem}
\usepackage{float}

\renewcommand{\sout}[1]{}
\renewcommand{\textcolor}[2]{#2}

\def\affNagoya{Nagoya University, Furocho, Chikusa, Nagoya 464-8602, Japan}
\def\affKyushu{Kyushu University, 744 Motooka, Nishi, Fukuoka 819-0395, Japan}
\def\affJAEA{Japan Atomic Energy Agency, 2-4 Shirakata, Tokai, Ibaraki 319-1195, Japan}
\def\affTokyoTech{Institute of Science Tokyo, Meguro, Tokyo 152-8551, Japan}
\def\affIbaraki{Ibaraki University, 2-1-1 Bunkyo, Mito, Ibaraki 310-8512, Japan}
\def\affRCNP{Osaka University, Ibaraki, Osaka 567-0047, Japan}
\def\affIndiana{Indiana University, Bloomington, Indiana 47401, USA}

\def\affKEK{High Energy Accelerator Research Organization, 1-1 Oho, Tsukuba, Ibaraki 305-0801, Japan}
\def\affTohoku{Tohoku University, 2-1-1 Katahira, Aoba, Sendai, 980-8576 Japan}
\def\affCross{Comprehensive Research Organization for Science and Society, Tokai, Ibaraki 319-1106, Japan}
\def\affSouthCarolina{University of South Carolina, Columbia, South Carolina 29208, USA}
\def\affTodai{The University of Tokyo, Kashiwa, Chiba 277-8581, Japan}
\begin{document}


\title{First Constraint on P-odd/T-odd Cross Section in Polarized Neutron Transmission through Transversely Polarized $^{139}$La}

\author[1]{Rintaro~Nakabe}
\affil{\affJAEA \email{nakabe.rintaro@jaea.go.jp}}
\author[2]{Clayton~J.~Auton}
\affil{\affIndiana}
\author[1]{Shunsuke~Endo}
\author[3]{Hiroyuki~Fujioka}
\affil{\affTokyoTech}
\author[4]{Vladimir~Gudkov}
\affil{\affSouthCarolina}
\author[5]{Katsuya~Hirota}
\affil{\affKEK}
\author[6]{Ikuo~Ide}
\affil{\affNagoya}
\author[5]{Takashi~Ino}
\author[7]{Motoyuki~Ishikado}
\affil{\affCross}
\author[1]{Wataru~Kambara}
\author[6,1]{Shiori~Kawamura}
\author[1]{Atsushi~Kimura}
\author[6]{Masaaki~Kitaguchi}
\author[1]{Ryuju~Kobayashi}
\affil{\affIbaraki}
\author[5]{Takahiro~Okamura}
\author[1,8]{Takayuki~Oku}
\author[6,1]{Takuya~Okudaira}
\author[6]{Mao~Okuizumi}
\author[2]{J.~G.~Otero~Munoz}
\author[7]{Joseph~D.~Parker}
\author[1]{Kenji~Sakai}
\author[9]{Tatsushi~Shima}
\affil{\affRCNP}
\author[6]{Hirohiko~M.~Shimizu}
\author[1]{Takenao~Shinohara}
\author[2]{William~M.~Snow}
\author[10,1]{Shusuke~Takada}
\affil{\affTohoku}
\author[1]{Ryuta~Takahashi}
\author[11]{Shingo~Takahashi}
\affil{\affTodai}
\author[1]{Yusuke~Tsuchikawa}
\author[12]{Tamaki~Yoshioka}
\affil{\affKyushu}

\begin{abstract}%
We report the first constraint on time-reversal invariance violating (TRIV) effects in polarized neutron transmission through a transversely polarized $^{139}$La target. We formulate the transmission asymmetry within the density matrix formalism, explicitly incorporating the forward scattering amplitude of $^{139}$La including tensor polarization terms up to third-rank. The formalism is applied to existing transmission data originally obtained to measure the spin-dependent cross section near the $0.75$~eV $p$-wave resonance. Since these data were not optimized for P-odd/T-odd observables, the attainable sensitivity is intrinsically limited; nevertheless, they provide a useful test of the formalism on real experimental data. No statistically significant TRIV signal is observed. By analyzing the global $\chi^2$ structure in the parameter space, we obtain an upper limit of $|W_T|<15~\mathrm{eV}$ at the 90\% confidence level. This corresponds to an upper limit on the resonance-averaged TRIV cross section of $|\Delta\sigma_{\not{T}\not{P}}|<8.3\times10^2~\mathrm{b}$. These results validate the present theoretical framework and provide guidance for future dedicated TRIV searches in polarized neutron transmission experiments.
\end{abstract}
\subjectindex{D00, D21, C03}

\maketitle

\section{Introduction}

 The search for time-reversal invariance violating (TRIV) effects offers an experimental approach to CP violation beyond the standard model of elementary particles via the CPT theorem. A novel experimental approach to search for TRIV effects in neutron spin propagation through polarized nuclear targets is being developed by the Neutron Optical and Time-Reversal Experiment (NOPTREX) collaboration as part of the J-PARC E99 program. A sensitive probe of TRIV effects is provided by measuring the transmission of low-energy polarized neutrons through a polarized nuclear target~\cite{Stodolsky1986,Kabir1982,Kabir1988}. The forward elastic scattering amplitude for such neutrons can be expressed as the sum of a spin-independent term and three spin-dependent terms:
\begin{equation}
    f=\alpha + \bm{\beta}\cdot\bm{\sigma}_n,\quad
 \bm{\beta}=\beta_{\hat{\bm{I}}}\hat{\bm{I}} + \beta_{\hat{\bm{k}}_n}\hat{\bm{k}}_n + \beta_{\hat{\bm{k}}_n\times \hat{\bm{I}}}\hat{\bm{k}}_n\times \hat{\bm{I}}
\label{eq_forward}
\end{equation}
where $\bm{\sigma}_n$ denotes the neutron spin operator, and $\hat{\bm{I}}$ and $\hat{\bm{k}}_n$ are unit vectors along the nuclear spin and neutron momentum, respectively. In particular, the $\beta_{\hat{\bm{k}}_n\times\hat{\bm{I}}}$ component is odd under both parity and time-reversal transformations, making it a clean signature of TRIV effects. Possible false asymmetries, commonly referred to as final-state interactions or nuclear recoil effects, can be experimentally controlled by suppressing and/or varying contamination from non-forward-scattered neutrons.

In neutron-induced compound nuclear processes, parity-violating (PV) effects as large as $10^{-2}$–$10^{-1}$ have been observed in $p$-wave resonances near $s$-wave tails, representing enhancements of up to six orders of magnitude compared to PV effects in nucleon-nucleon interactions~\cite{Alfimenkov1983,Masuda1989,Yuan1991,Shimizu1993,Mitchell2001}. This enhancement can be interpreted as a statistical amplification of many small contributions, arising from the large number of degrees of freedom in compound nuclei~\cite{Bunakov1982,Bunakov1983,Gudkov1992,Bowman2014,Sushkov1982}. A similar mechanism is expected to enhance TRIV effects as well~\cite{Bunakov1982,Gudkov1992,Bowman2014}. A quantitative connection between TRIV and PV cross sections is established through a spin-dependent factor $\kappa(J)$, which depends on the compound-nuclear spin and the partial neutron widths of $p$-wave resonances~\cite{Bunakov1983,Gudkov1990}. The TRIV and PV cross sections, $\Delta \sigma_{\not{T}\not{P}}$ and $\Delta \sigma_{\not{P}}$, are related by 
\begin{equation}
   \Delta \sigma_{\not{T}\not{P}} = \kappa(J) \frac{W_T}{W} \Delta \sigma_{\not{P}},
\end{equation}
where $W_T$ and $W$ denote the TRIV and PV matrix elements of the nucleon-nucleon interaction, respectively~\cite{Bunakov1982}. For a compound nuclear spin $J = I + 1/2$, the value of $\kappa(J)$ is given by~\cite{Gudkov2018}
\begin{equation}
\kappa(J) = \frac{I}{I+1} \left(1 + \frac{1}{2} \sqrt{\frac{2I + 3}{I}} \frac{y}{x} \right),
\label{eq_kappa}
\end{equation}
where $I$ is the spin of the target nucleus. Here, $x$ and $y$ are defined by $x^2=\Gamma_{p,j=1/2}^n/\Gamma_p^n$ and $y^2=\Gamma_{p,j=3/2}^n/\Gamma_p^n$~\cite{Flambaum1985}, so that $x^2+y^2=1$. A corresponding mixing angle $\phi$ is introduced through $x=\cos\phi$ and $y=\sin\phi$.

The sensitivity to TRIV effects is proportional to the product $\Delta \sigma_{\not{P}} \kappa(J)$. Previous measurements have reported a longitudinal asymmetry of $9.55 \pm 0.35\%$ in the neutron absorption reaction of $^{139}$La at a $0.75\,\mathrm{eV}$ $p$-wave resonance~\cite{Yuan1991}. Moreover, the value of $\kappa(J)$ for $^{139}$La has been experimentally determined to be $0.59 \pm 0.05$~\cite{Nakabe2024}. These features make $^{139}$La a particularly promising candidate for TRIV searches in neutron transmission experiments. However, several challenges remain to be addressed in the high-sensitivity search for TRIV effects in transmission experiments. One is the technical difficulty of controlling neutron and nuclear spins in transmission geometry, and the other is the development of a rigorous theoretical framework.

In this study, we derive a constraint on TRIV effects by applying the density matrix formalism for neutron spin propagation to the transmission data reported in Ref.~\cite{Okudaira2024}, incorporating the forward scattering amplitude of $^{139}$La expanded up to third-rank tensor polarizations. The data employed in this analysis were not optimized for P-odd/T-odd observables but were originally obtained from an experiment aimed at measuring the spin-dependent cross section at the $p$-wave resonance. Accordingly, it is important to emphasize that the sensitivity to $W_T/W$ is intrinsically orders of magnitude lower than that achievable in future dedicated experiments. Preliminary results were reported in~\cite{NakabeProc2024}. This paper provides a complete theoretical formulation and a quantitative reanalysis of the experimental data, extending the previous work. It demonstrates the applicability of the framework to real data and establishes a methodology for extracting constraints on TRIV observables from neutron transmission experiments.

\section{Spin observables in transmission through polarized $^{139}$La}
\subsection{Observed asymmetry in neutron transmission}

In this section, we present only the expressions required for the analysis in Sec.~3, specialized to the experimental configuration. The general formulation is provided in Appendices~\ref{appandix_A} and~\ref{Appndix_B}. 

The theoretical asymmetry (or analyzing power) $A_I(\vartheta_{kI})$, where $\vartheta_{kI}$ denotes the angle between $\hat{\bm{k}}_n$ and $\hat{\bm{I}}$, is defined as
\begin{equation}
    A_I(\vartheta_{kI}) = \frac{1}{p_n}\frac{N_- - N_+}{N_+ + N_-}
    = -\frac{2 \left[ \mathrm{Re}(A^*B) + \mathrm{Im}(C^*D) \right]}{|A|^2 + |B|^2 + |C|^2 + |D|^2},
    \label{eq_anapow}
\end{equation}
where $N_+$ and $N_-$ represent the transmitted neutron counts (or expectation values) for the spin-up and spin-down initial states. The neutron polarization is denoted by $p_n$ with its spin aligned parallel to the nuclear spin. The dimensionless quantities $A$, $B$, $C$, and $D$ are defined as
\begin{equation}
\begin{aligned}
A = e^{iZ\alpha}\cos(Z\tilde{\beta}),\quad
(B,\,D,\,C) = iZ e^{iZ\alpha}\frac{\sin(Z\tilde{\beta})}{Z\tilde{\beta}}
\left(
\beta_{\hat{\bm{I}}}-\mu_{\rm eff}B_{\rm ext},\,
\beta_{\hat{\bm{k}}_n\times\hat{\bm{I}}},\,
\beta_{\hat{\bm{k}}_n}
\right),
\end{aligned}
\label{eq_ABCD}
\end{equation}
where the ordering $(B, D, C)$ corresponds to the components $(\tilde{\beta}_x, \tilde{\beta}_y, \tilde{\beta}_z)$. Here, $Z = 2\pi \rho z / k_n$ denotes the wavenumber-weighted column density, where $\rho$ is the number density of target nuclei and $z$ is the thickness of the target. The spin-correlation amplitude $\tilde{\beta}$ is expressed using Eq.~\ref{eq_forward} as
\begin{equation}
\tilde{\beta}=\sqrt{\left(\beta_{\hat{\bm{I}}}-\mu_{\rm eff}B_{\rm ext}\right)^2+\beta_{\hat{\bm{k}}_n\times\hat{\bm{I}}}^2+\beta_{\hat{\bm{k}}_n}^2},
\label{eq_beta_expl}
\end{equation}
where $\bm{B}_{\rm ext}$ is the external magnetic field, and $\mu_{\rm eff} \equiv - m_n\mu_n / (2\pi\hbar^2 \rho)$ is the effective magnetic moment in the target, defined in terms of the neutron mass $m_n$ and the neutron magnetic moment $\mu_n$. To clarify the contribution of each term to the asymmetry, we rewrite the interference terms appearing in Eq.~\ref{eq_anapow} as
\begin{equation}
A^*B = \frac{e^{-2\,\mathrm{Im}(Z\alpha)}}{2|\tilde{\beta}|^2}
\left[ i\zeta_s - \eta_s \right]
\tilde{\beta}^* \left(\beta_{\hat{\bm{I}}}-\mu_{\rm eff}B_{\rm ext}\right),\quad
C^*D = \frac{e^{-2\,\mathrm{Im}(Z\alpha)}}{2|\tilde{\beta}|^2}
\left[ i\zeta_c + \eta_c \right]
\beta_{\hat{\bm{k}}_n}^* \beta_{\hat{\bm{k}}_n\times\hat{\bm{I}}},
\label{eq_AXi}
\end{equation}
and the denominator is given by
\begin{equation}
|A|^2 + |B|^2 + |C|^2 + |D|^2
= \frac{e^{-2\,\mathrm{Im}(Z\alpha)}}{2}
\left[
\zeta_c + \eta_c - (\zeta_c-\eta_c)\frac{|\tilde{\beta}_x|^2+|\tilde{\beta}_y|^2+|\tilde{\beta}_z|^2}{|\tilde{\beta}|^2}
\right],
\label{eq_ABCD_norm}
\end{equation}
where $\zeta_c \equiv \cos(2\,\mathrm{Re}(Z\tilde{\beta}))$, $\zeta_s \equiv \sin(2\,\mathrm{Re}(Z\tilde{\beta}))$, $\eta_c \equiv \cosh(2\,\mathrm{Im}(Z\tilde{\beta}))$, and $\eta_s \equiv \sinh(2\,\mathrm{Im}(Z\tilde{\beta}))$. This expression indicates that oscillatory and absorptive behaviors are associated with $\mathrm{Re}(Z\tilde{\beta})$ and $\mathrm{Im}(Z\tilde{\beta})$, respectively, although both contributions are mixed in the observable asymmetry. This formulation provides a direct connection between the measured asymmetry and the underlying spin-dependent forward scattering amplitudes. To proceed further, the explicit form of the spin-correlation amplitudes $\alpha$ and $\tilde{\bm{\beta}}$ must be specified for the target nucleus under consideration.

\subsection{Spin-correlation amplitudes for $^{139}$La}

The forward scattering amplitude is typically described using four spin-correlation terms. For the $^{139}$La nucleus with spin $I = 7/2$, additional contributions from higher-rank spherical tensors become relevant. In the low-energy neutron regime, higher orbital angular momenta ($l > 1$) can be neglected, and the expansion is truncated at tensor rank $q \leq 3$. The forward scattering amplitude for $^{139}$La, incorporating tensor polarizations up to rank 3, is given in Ref.~\cite{Gudkov2020} as
\begin{equation}
\begin{aligned}
\alpha &= A' + P_1 H' \cos\vartheta_{kI}
+ P_2 E' \left(\cos^2\vartheta_{kI} - \frac{1}{3}\right), \\
\beta_{\hat{\bm{I}}} &= P_1 B' + P_2 F' \cos\vartheta_{kI}
+ P_3 \frac{B_3'}{3} \left(\cos^2\vartheta_{kI} - 1\right), \\
\beta_{\hat{\bm{k}}_n\times \hat{\bm{I}}} &= P_1 D',\\
\beta_{\hat{\bm{k}}_n} &= C' + P_1 K' \cos\vartheta_{kI}-\frac{1}{3} P_2 F' + \frac{2}{3} P_3 B_3' \cos\vartheta_{kI},
\label{eq_coefficients}
\end{aligned}
\end{equation}
where $P_1$, $P_2$, and $P_3$ are the tensor polarization components of rank 1, 2, and 3, respectively. Additional coefficients, $H'$, $K'$, $E'$, $F'$, and $B_3'$, associated with higher-rank tensor contributions and the angular dependence on $\cos\vartheta_{kI}$, appear in the scattering amplitude. The primed coefficients are calculated within the two-resonance approximation including $s$- and $p$-wave states. Although these expressions are based on Ref.~\cite{Gudkov2020}, we rederive them here to ensure consistency and clarity, and present the resulting forms explicitly. The coefficients are given by
\begin{equation}
	\begin{aligned}
		A'=& -\frac{1}{32k_n}\left(9t_{s0}+7t_{s1}+9t_{p}\right) + \frac{i}{16}\left(9t_{s0}a_{s0}+7t_{s1}a_{s1}\right)-\frac{1}{16}\left(9a_{s0}+7a_{s1}\right)+\frac{ik_n}{16}\left(9a_{s0}^2+7a_{s1}^2\right)\\              
        B'=&-\frac{1}{32k_n}\left(7t_{s0}-7t_{s1}+t_{p}\left(-7x^2-2\sqrt{35}xy+\frac{2}{5}y^2 \right)\right)\\ &\quad+\frac{7i}{16}\left(t_{s0}a_{s0}-t_{s1}a_{s1}\right)-\frac{7}{16}\left(a_{s0}-a_{s1}\right)+\frac{7ik_n}{16}\left(a_{s0}^2-a_{s1}^2\right),\\
        C'=&-\frac{9xW}{16k_n}t_{s0,p},\quad D'=-\frac{W_{T}}{16k_n}\left( 7x+\sqrt{35}y \right)t_{s0,p},\quad H'=-\frac{W}{16 k_n}\left(7x-2\sqrt{35}y\right) t_{s0,p}\\                
        K'=&-\frac{1}{16 k_n} \left(7 x^2-\sqrt{35} x y -\frac{1}{10}y^2\right)t_{p},\\
        E'=&\frac{9}{320 k_n} \left(4 \sqrt{35} x y- 13 y^2\right)t_p,\quad
        F'=\frac{9 \sqrt{35}yW}{80 k_n} t_{s0,p},\quad B_3'=\frac{81y^2}{320 k_n} t_p,
	\end{aligned}
\label{eq_FSA_La}	
\end{equation}
where $t_{s0}$, $t_{s1}$, and $t_{p}$ are Breit-Wigner amplitudes for the negative $s$-wave, $s$-wave at the $72$~eV, and $p$-wave resonances, written as
\begin{equation}
    t_{s0}=\frac{\Gamma^n_{s0}}{E-E_{s0}+i\Gamma_{s0}/2},\quad t_{s1}=\frac{\Gamma^n_{s1}}{E-E_{s1}+i\Gamma_{s1}/2},\quad t_{p}=\frac{\Gamma^n_{p}}{E-E_{p}+i\Gamma_{p}/2},
\end{equation}
and $t_{s0,p}$ is
\begin{equation}
    t_{s0,p}=\frac{\sqrt{\Gamma^n_{s0}} \sqrt{\Gamma^n_{p}}}{(E-E_p+i\Gamma_p/2)(E-E_{s0}+i\Gamma_{s0}/2)}.
\end{equation}
The resonance parameters $E_K$, $\Gamma^n_K$, and $\Gamma_K$ are the resonance energy, neutron width, and total width of $K$th resonance. The potential scattering lengths of the $s_0$- and $s_1$-waves are $a_{s0}$ and $a_{s1}$, respectively. Since the spin coupling schemes differ between Eq.~\ref{eq_kappa} and Refs.~\cite{Gudkov2020,Gudkov2018}, a conversion is applied to relate the two conventions, $x_S=-\frac{1}{2\sqrt{3}}(\sqrt{7}x+\sqrt{5}y)$ and $y_S=\frac{1}{2\sqrt{3}}(\sqrt{5}x-\sqrt{7}y)$. Thus, the observed asymmetry in Eq.~\ref{eq_anapow} can be evaluated numerically by substituting the resonance parameters into Eq.~\ref{eq_FSA_La}.

\section{Evaluation of the TRIV upper limit}
\subsection{Physical interpretation of the previous experiment}

In the present analysis, we reanalyze the experimental asymmetry reported in Ref.~\cite{Okudaira2024} with the aim of extracting constraints on the TRIV interaction parameter $W_T$. Although the original experiment focused on the measurement of the spin-dependent cross section at the $p$-wave resonance using polarized neutrons transmitted through a transversely polarized $^{139}$La target, the present study interprets the same dataset within the extended theoretical framework developed in the previous section, which explicitly includes the TRIV contribution through the forward scattering amplitude. In this experimental configuration, the scalar triple product $\bm{\sigma}_n\cdot(\hat{\bm{k}}_n\times\hat{\bm{I}})$ vanishes because the neutron spin is aligned along $\hat{\bm{I}}$. Nevertheless, a TRIV contribution can still arise through the structure of neutron spin propagation in the polarized target. In the absence of TRIV effects, the dominant spin-dependent amplitude aligns the effective spin-quantization axis approximately along $\hat{\bm{I}}$, whereas the TRIV amplitude $D'$ introduces a transverse component. As a consequence, the absorption associated with ${\rm Im}D'$ is projected onto the measurement basis defined along $\hat{\bm{I}}$, leading to a contribution proportional to ${\rm Re}D'\,{\rm Im}D'$. Therefore, the present analysis does not introduce new experimental data, but instead provides a reinterpretation of existing measurements to derive constraints on TRIV observables.

\subsection{Experimental configuration}

    \begin{figure}[h]
        \centering
        \includegraphics[scale=0.26]{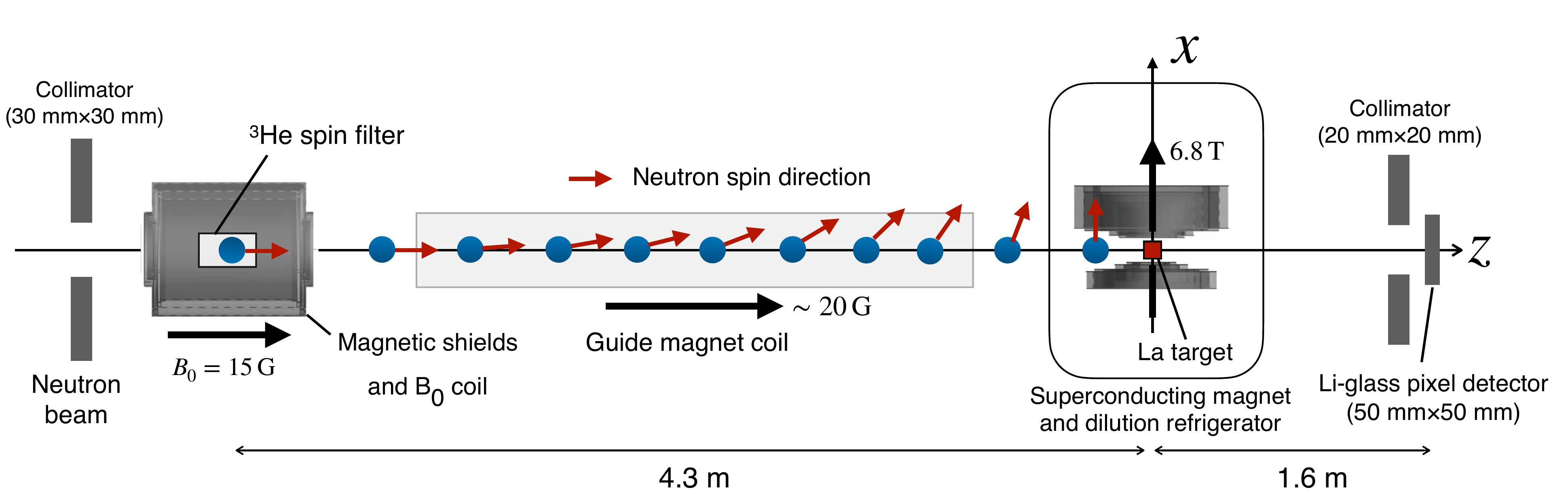}
        \caption{Schematic layout of the polarized-neutron transmission experiment through a transversely polarized $^{139}$La target. The incident pulsed neutron beam is polarized by a $^3$He spin filter under a longitudinal holding field ($B_0=15$ G). The neutron spin is transported downstream by a guide field ($\sim 20$ G), during which the spin direction gradually follows the local magnetic field. Upon entering the superconducting magnet, a strong transverse field of 6.8 T aligns the neutron spin along the $x$ axis, where the $^{139}$La target is polarized. The neutron momentum defines the $z$ axis.}
        \label{fig_exp}
    \end{figure}

The experiment was performed using a pulsed neutron beam at the RADEN beamline~\cite{RADEN}, located at the Material and Life Science Experimental Facility (MLF)~\cite{Nakajima2017} at the Japan Proton Accelerator Research Complex (J-PARC). The repetition rate of the proton beam was 25~Hz, and the beam power during the experiment was 750~kW. A schematic layout of the experimental setup, in which a polarized neutron beam is transmitted through a polarized target, is shown in Fig.~\ref{fig_exp}. The neutron beam, moderated by liquid hydrogen, passed through a 1~mm-thick cadmium filter located 8~m downstream of the moderator surface (defined as 0~m), thereby removing thermal neutrons that would contribute to the heat load on the lanthanum target and to unwanted background counts. The incident neutron beam was collimated by a $30\times30$~mm$^2$ aperture at 18.7~m and subsequently polarized by a $^3$He spin filter device~\cite{He3Okudaira}. The polarized beam was then incident on a lanthanum metal target placed in a transverse magnetic field of $|\bm{B}_{\rm ext}|=6.8~{\rm T}$. The neutron spin was transported adiabatically along the combined magnetic field formed by the guide coil applied along the beam axis and the external magnetic field at the target. The neutron spin flip was achieved by reversing the $^3$He polarization using an adiabatic fast-passage (AFP) NMR technique with an RF coil integrated into the $^3$He spin filter device. The loss of the $^3$He polarization associated with each spin flip was $4\times10^{-5}$. The neutron polarization $p_n$ was evaluated as a function of neutron energy from transmission measurements with polarized and unpolarized $^3$He. The averaged neutron polarization at the $^{139}$La $p$-wave resonance energy of 0.75~eV was $36.1\pm0.5\%$. The target consisted of a cubic lanthanum metal with a side length of 2.0~cm, which was thermally connected to a copper cold head of a dilution refrigerator. A temperature of 67~mK was recorded by a RuO$_2$ thermometer attached to the cold head; however, thermal gradients due to heat inflow through conduction and radiation, as well as the Kapitza resistance at interfaces, must be taken into account. Therefore, the target temperature was evaluated using the measured spin-dependent cross section at the 2.99~eV $s$-wave resonance of the impurity isotope $^{138}$La contained in the lanthanum metal. From the nuclear magnetic moment and spin of $^{138}$La, the target temperature was determined to be $75.7^{+10.2}_{-8.9}$~mK. Under this thermal-equilibrium condition, the tensor polarization components were evaluated as $P_1=3.9\pm0.5\%$, $P_2=0.1^{+0.03}_{-0.02}\%$, and $P_3=(2.1\pm1.0)\times10^{-3}\%$. Neutrons transmitted through the target passed through a $20\times20$~mm$^2$ collimator located 1.6~m downstream of the target and were counted using a PMT-based Li-glass scintillation detector~\cite{Liglass2015}. The beam divergence, corresponding to $\cos\vartheta_{kI}$ in Eq.~\ref{eq_coefficients}, was below $2\times10^{-3}$ when considering the maximum angular spread along the flight path.  

\subsection{Observable and dominant spin-correlation terms}

The transmitted neutron counts for parallel and antiparallel spin configurations between the neutron polarization and the target nuclear polarization were measured as $N_+$ and $N_-$, respectively. During the asymmetry measurement, the neutron spin was flipped every 30 minutes by reversing the $^3$He polarization. The experimental asymmetry corrected for the neutron polarization was quantified in terms of the analyzing power, defined as
\begin{equation}
        \left(A_I(\vartheta_{kI})\right)_{\rm exp}=\frac{1}{p_n}\frac{N_--N_+}{N_-+N_+},
        \label{eq_anapow_obs}
\end{equation}
where the transmitted neutrons were measured using the time-of-flight (TOF) method. The relation between the incident neutron energy and TOF is given by 
$E\,{\rm (eV)} = 5.227\times10^{-3}\,\left({\rm LOF\,(m)}/{\rm TOF\,(ms)}\right)^2$, where the length of flight (LOF) was determined to be 24.713~m. We first identify the dominant spin-correlation terms in Eq.~\ref{eq_anapow}. As shown in Eqs.~\ref{eq_AXi} and \ref{eq_ABCD_norm}, when $(|\tilde{\beta}_x|^2+|\tilde{\beta}_y|^2+|\tilde{\beta}_z|^2)/|\tilde{\beta}|^2 \simeq 1$, the second‐order contribution of $\tilde{\beta}$ can be neglected, and only the first-order term needs to be retained. Since $|D'| \ll |B'|$, it follows that $|\tilde{\beta}_{y}|^2 \ll |\tilde{\beta}_{x}|^2$. We obtained a numerical estimate around the $p$-wave resonance as $|\tilde{\beta}_{z}|^2/(|\tilde{\beta}_x|^2+|\tilde{\beta}_y|^2+|\tilde{\beta}_z|^2)=4\times10^{-11}$. 
\begin{table} [h]
  \centering
  \caption{Resonance parameters of $^{139}$La taken from Ref.~\cite{Endo2023}.}
  \label{table_nuclear_par} 
  \begin{tabular}{lcccccc} 
    \hline\hline
	$K$&
	$E_{K}\rm\,(eV)$&
	$J_{K}$&
	$l_{K}$&
	$\Gamma_{K}^\gamma \rm \,(meV)$&
	$g_{K}\Gamma_{K}^n \rm \,(meV)$\\
    \hline
    	$0$ & $-38.8\pm0.4$ & $4$ & 0 & $60.3\pm0.5$ & $346\pm10$ &  $s_{0}$\\
	$1$ & $0.750\pm0.001$ & $4$ & 1 & $41.6\pm0.9$ & $(3.67\pm0.05)\times10^{-5}$  & $p$\\
	$2$ & $72.30\pm0.01$ & $3$ & 0 & $64.1\pm3.0$ & $13.1\pm0.7$ &  $s_{1}$\\
\hline\hline
  \end{tabular}
  \normalsize 
\end{table}
In evaluating the spin-correlation terms $B', C', K', F'$, and $B_3'$, we chose the mixing angle $\phi$ to maximize each term individually. The P-odd matrix element was assumed to be $W\sim1\,{\rm meV}$ in Refs.~\cite{Frankle1991,Bowman1990,Fadeev2019}, and all resonance parameters and physical constants were taken to be the same as in Ref.~\cite{Okudaira2024} as listed in Tab.~\ref{table_nuclear_par}. The second term, ${\rm Im}C^{*}D$, vanishes because the resonance $t_{s0,p}$ is common to both, giving ${\rm Im}C'^{*}D' = 0$. Moreover, the contributions from spin-correlation terms other than the leading ones, as demonstrated below, are negligible. Under these approximations, we obtain $\mathrm{Re}A^{*}B  \propto\eta_s$ and $|A|^2+|B|^2+|C|^2+|D|^2\propto\eta_c$. Accordingly, Eq.~\ref{eq_anapow} reduces to
\begin{equation}
    A_I(\vartheta_{kI}) \simeq \tanh(2\mathrm{Im}(Z\tilde{\beta})).
    \label{eq_Ax_theo}
\end{equation}
To identify the dominant contribution to the first–order ${\rm Im}(Z\tilde{\beta})$, we examined the residual of the ratio between computations with and without each spin correlation term. 
\begin{table} [h]
  \centering
  \caption{Residual from unity of the ratio computed by excluding a specific correlation term included in $\mathrm{Im}\tilde{\beta}$. Since the real part of the resonance enters in differential form, $\mathrm{Im}\tilde{\beta}$ was evaluated using the value integrated over $\pm 3\Gamma_p$ around the $p$-wave resonance.}
  \label{table_beta} 
  \begin{tabular}{c|cccccc} 
    \hline\hline
    $\epsilon$ & $P_1B'=0$ & $\mu_{\rm eff}B_{\rm ext}=0$ & $C'=0$ & $P_2F'=0$ & $P_1K'(\hat{\bm{k}}_n\cdot \hat{\bm{I}})=0$ & $P_3B_3'=0$\\
    \hline
    $1-\frac{{\rm Im}\tilde{\beta}(\epsilon)}{{\rm Im}\tilde{\beta}}$ & $1$ & $0.9$ & $2\times 10^{-6}$ & $1\times10^{-4}$ & $2\times10^{-8}$ & $2\times10^{-4}$\\
    \hline\hline
  \end{tabular}
  \normalsize 
\end{table}
Table~\ref{table_beta} shows that contributions proportional to tensor polarizations of second and third rank, beam divergence, and the P-odd term $C'$ are all below $10^{-4}$. Given that the spin-dependent cross section at the $p$-wave resonance has a statistical relative uncertainty of $0.3$~\cite{Okudaira2024}, we can therefore approximate
\begin{equation}
    \tilde{\beta} \simeq \sqrt{(P_1B'-\mu_{\rm eff}B_{\rm ext})^2+(P_1D')^2}.
    \label{eq_beta_theo}
\end{equation}
To clarify the contribution of the $D'$ term, we expand ${\rm Im}\tilde{\beta}$. In the regime where the external magnetic field dominates, it can be approximated as
    \begin{equation}
        {\rm Im}\tilde{\beta} \simeq P_1{\rm Im}B' + \frac{P_1^2{\rm Re}D'{\rm Im}D'}{\mu_{\rm eff}B_{\rm ext}}. 
        \label{eq_Imbeta_appro}
    \end{equation}
This shows that the observable TRIV contribution is suppressed by the inverse of the external magnetic field and scales with the square of the nuclear polarization.

\subsection{Evaluation of the TRIV upper limit}

\begin{figure}[h]
        \centering
        \includegraphics[width=\linewidth]{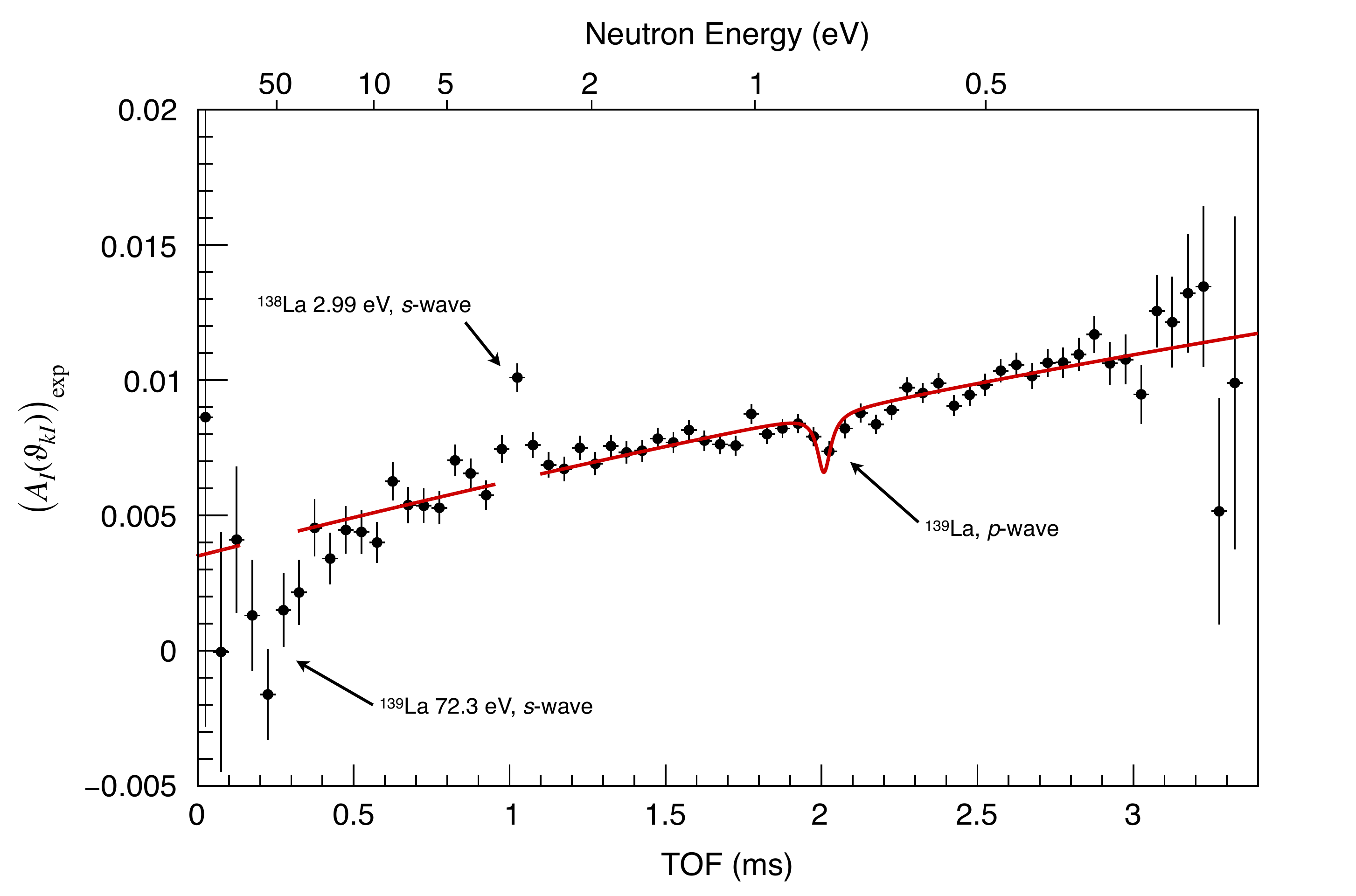}
        \caption{Measured asymmetry $\left(A_I(\vartheta_{kI})\right)_{\rm exp}$ as a function of neutron time-of-flight (TOF). The red solid line represents the best-fit function consisting of the $p$-wave resonance asymmetry $\left(A_I(\vartheta_{kI})\right)_p$ and a quadratic background term describing the smoothly varying contribution from the $s$-wave resonances. The regions around the $s_1$ resonance at 72.3~eV and the $^{138}$La $s$-wave resonance at 2.99~eV were excluded from the fit. The 2.99~eV resonance originates from a 0.09\% $^{138}$La impurity, for which $J=11/2$ was assigned in Ref.~\cite{Alfimenkov1993}.}
        \label{fig_w_v_Fit}
    \end{figure}

The global structure of the measured asymmetry $\left(A_I(\vartheta_{kI})\right)_{\rm exp}$, shown in Fig.~\ref{fig_w_v_Fit}, is primarily governed by the $s_0$ and $s_1$ wave resonances. 
These contributions originate from the independent potential scattering lengths $a_{s0}$ and $a_{s1}$ included in the $B'$ term in the forward scattering amplitude. Although the values of $a_{s0}$ and $a_{s1}$ can be taken from previous studies~\cite{Endo2023}, the uncertainties in these parameters may introduce a significant discrepancy in the overall scale of the asymmetry predicted by the theoretical expression. Since the primary objective of this study is to evaluate the TRIV effect in the vicinity of the $p$-wave resonance, the contributions from the $s$-wave resonances were treated as a smoothly varying background rather than modeled in detail. Specifically, Gaussian functions were fitted to the $s_1$-wave resonance, the $^{138}$La $s$-wave resonance, and the $p$-wave resonance, and the regions within $\pm$FWHM around each resonance peak center were excluded from the fit. 
The remaining regions were then fitted with a quadratic function to describe the $s$-wave background. The fit yielded $\chi^2/{\rm dof}=0.986$, confirming that the background structure is adequately reproduced within the statistical uncertainties. The asymmetry arising solely from the forward scattering amplitude coefficients of the $p$-wave resonance is written as $\left(A_I(\vartheta_{kI})\right)_p$. Using Eqs.~\ref{eq_Ax_theo} and~\ref{eq_beta_theo}, this can be written as
    \begin{equation}
    \begin{array}{c}
        \left(A_I(\vartheta_{kI})\right)_p({\rm TOF};W_T,\phi,E_p) = \tanh\left(2{\rm Im}(Z\tilde{\beta}_p)\right),\\[1ex]
        \tilde{\beta}_p = \sqrt{\left(P_1B'_p({\rm TOF};\phi,E_p) - \mu_{\rm eff}B_{\rm ext}\right)^2 + \left(P_1D'({\rm TOF};W_T,\phi,E_p)\right)^2},
    \end{array}
    \end{equation}
where $B_p' = -\frac{1}{32k_n} t_p \left(-7x^2 - 2\sqrt{35}xy + \frac{2}{5}y^2 \right)$, and the product of the atomic number density and the target thickness is $\rho z=0.053\,{\rm b^{-1}}$. The measured asymmetry 
$\left(A_I(\vartheta_{kI})\right)_{\rm exp}$ was fitted over the range $0\,{\rm ms}<{\rm TOF}<3.4\,{\rm ms}$ using the following fitting function:
    \begin{equation}
        \left(A_I(\vartheta_{kI})\right)_{\rm fit}({\rm TOF};W_T,\phi,E_p,\mathbf{c})=\sum_{i=0,1,2}c_i({\rm TOF})^i + \left(A_I(\vartheta_{kI})\right)_p({\rm TOF};W_T,\phi,E_p).
    \end{equation}
The free parameters were the three coefficients of the quadratic background function, $c_i\,(i=0,1,2)$, together with the physical parameters $W_T$, $\phi$, and $E_p$, giving a total of six parameters. The result is shown in Fig.~\ref{fig_w_v_Fit}. The $s_1$ resonance at $72.3~{\rm eV}$ and the $^{138}$La $s$-wave resonance at $2.99~{\rm eV}$ were excluded from the fit region. The fit yielded $\chi^2=54.8$ for ${\rm dof}=56$, corresponding to $\chi^2/{\rm dof} \approx 0.98$, indicating that the fit describes the data well. Since the physical solution for the mixing angle $\phi$ is known to lie in the second quadrant from a previous study~\cite{Nakabe2024}, the fit was performed with the constraint $90^\circ<\phi<180^\circ$. The best-fit values were $\phi=(163.7\pm3.4)^\circ$, $W_T=0.02\pm12\,{\rm eV}$, and $E_p=0.792\pm0.006\,{\rm eV}$, where the quoted uncertainties correspond to $1\sigma$ (68.3\% confidence level).

\subsection{Discussion}

We now discuss the implications of the fit results and the structure of the $\chi^2$ parameter space. The fitted value of $\phi$ is consistent with the previously reported value of $\phi = 164\pm4^\circ$ in Ref.~\cite{Nakabe2024}. The correlation coefficients between the fitted parameters were $2\times10^{-3}$ for $(\phi, W_T)$, $0.5$ for $(\phi, E_p)$, and $5\times10^{-2}$ for $(W_T, E_p)$. These results indicate that $W_T$ has only weak correlations with the other parameters and is determined relatively independently. However, to evaluate the confidence interval of $W_T$, it is necessary to examine not only the parameter correlations near the minimum but also the global structure of the $\chi^2$ space. For this purpose, the parameter space was analyzed using $\chi^2$ contour maps and profile $\chi^2$. 

The $\chi^2$ contour maps were evaluated in the two parameter planes, $\phi-W_T$ and $W_T-E_p$. For each parameter plane, the profile $\chi^2$ is defined by minimizing $\chi^2$ with respect to the remaining parameters as $\chi^2_{\rm prof}(\phi,W_T)=\min_{E_p,\mathbf{c}}\chi^2(W_T,\phi,E_p,\mathbf{c})$, and $\chi^2_{\rm prof}(W_T,E_p)=\min_{\phi,\mathbf{c}}\chi^2(W_T,\phi,E_p,\mathbf{c})$. 
    \begin{figure}[h]
        \centering
        \includegraphics[width=1.02\linewidth]{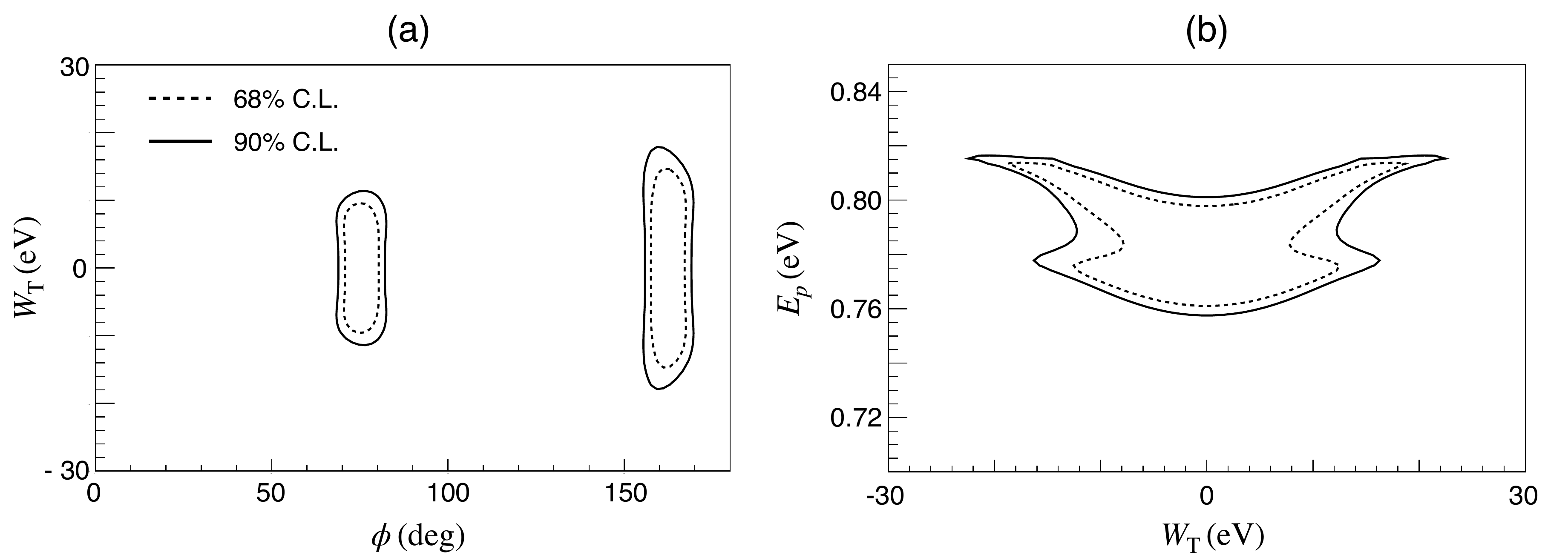}
        \caption{$\chi^2$ contour maps in the two-parameter planes (a) $(\phi, W_T)$ and (b) $(W_T, E_p)$. For each point in the plane, the profile $\chi^2$ value is obtained by minimizing $\chi^2$ with respect to the remaining parameters. The inner and outer contours correspond to $\Delta\chi^2=2.30$ and $4.61$, representing the 68\% and 90\% confidence levels for two degrees of freedom. Two local minima appear in the $\phi$ direction within the range $0^\circ < \phi < 180^\circ$. Since the structure is repeated in the range $180^\circ < \phi < 360^\circ$, this corresponds to the four-fold solutions reported in Ref.~\cite{Okudaira2024} when extended to the full angular range. In all local solutions, $W_T$ remains consistent with zero. }
        \label{fig:chi2map}
    \end{figure}
The resulting contour maps are shown in Fig.~\ref{fig:chi2map}. The contours correspond to $\Delta\chi^2=2.30$ and $4.61$ for two degrees of freedom, representing the 68\% and 90\% confidence levels, respectively. In the $\phi$--$W_T$ plane, two local minima appear with respect to $\phi$ within the range $0^\circ < \phi < 180^\circ$. Owing to the periodicity of the solution, the same structure is repeated in the range $180^\circ < \phi < 360^\circ$, leading to the four-fold solutions reported in Ref.~\cite{Okudaira2024}. In addition, $W_T$ is consistent with zero for all local solutions. In the $W_T-E_p$ plane, the $\chi^2$ surface exhibits a non-elliptic, multimodal structure rather than a simple valley minimum. In particular, two low-$\chi^2$ regions appear around $E_p\sim0.76$ and $0.80~{\rm eV}$, indicating that different combinations of $W_T$ and $E_p$ can reproduce the shape of the observed $p$-wave asymmetry comparably well. To quantify these features of the parameter space further, one-dimensional $\chi^2$ profiles were evaluated.

The one-dimensional $\Delta\chi^2$ profiles are defined as $\Delta\chi^2(W_T)=\min_{\phi,E_p,\mathbf{c}}\chi^2(W_T,\phi,E_p,\mathbf{c})-\chi^2_{\min}$ and $\Delta\chi^2(E_p)=\min_{\phi,W_T,\mathbf{c}}\chi^2(W_T,\phi,E_p,\mathbf{c})-\chi^2_{\min}$. Here, $\chi^2_{\min}$ denotes the minimum $\chi^2$ value in the full parameter space $(W_T, \phi, E_p, \mathbf{c})$. 
    \begin{figure}[]
        \centering
        \includegraphics[width=1.02\linewidth]{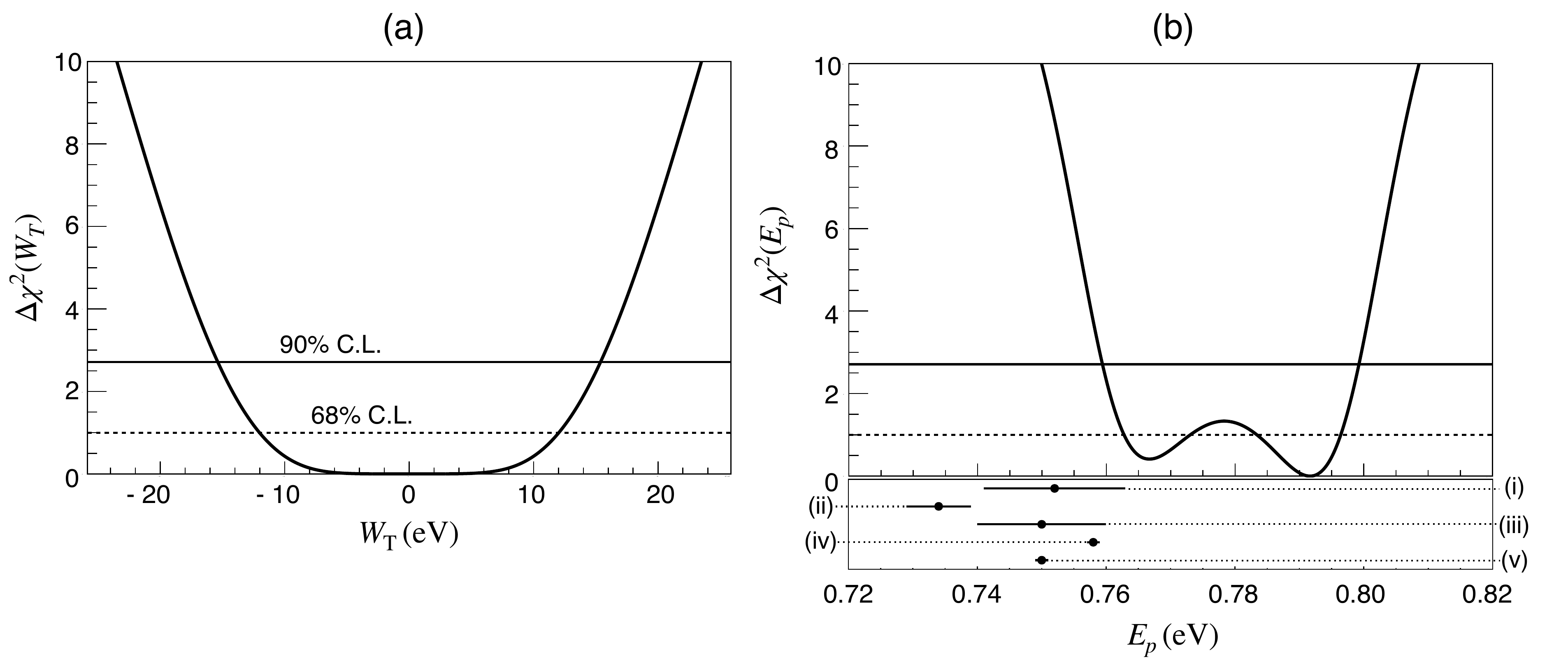}
        \caption{One-dimensional profile $\Delta\chi^2$ distributions for (a) $W_T$ and (b) $E_p$. For each value of the parameter of interest, the $\chi^2$ is minimized with respect to the remaining parameters. The horizontal lines correspond to $\Delta\chi^2=1.0$ and $2.71$, representing the 68\% and 90\% confidence levels for one degree of freedom. Panel (a) shows that the minimum occurs near $W_T \approx 0$, indicating no statistically significant TRIV signal. The intersections with $\Delta\chi^2=2.71$ give the upper limit $|W_T| < 15\,{\rm eV}$ at the 90\% confidence level. Panel (b) exhibits two local minima around $E_p \approx 0.766\,{\rm eV}$ and $E_p \approx 0.792\,{\rm eV}$, reflecting a multimodal structure of the $\chi^2$ surface. The lower panel shows literature values of $E_p$ with horizontal error bars for comparison: (i) Harvey \textit{et al.} (1959)~\cite{Harvey_1959}; (ii) Shwe \textit{et al.} (1967)~\cite{Shwe_1967}; (iii) Alfimenkov \textit{et al.} (1983)~\cite{Alfimenkov1983}; (iv) Terlizzi \textit{et al.} (2007)~\cite{Terlizzi2007}; (v) Endo \textit{et al.} (2023)~\cite{Endo2023}.}
        \label{fig:chi2prof}
    \end{figure}
The profile $\Delta\chi^2(W_T)$ is shown in Fig.~\ref{fig:chi2prof}(a). The profile has a minimum at $W_T = 0.02 \pm 12\,{\rm eV}$, which is consistent with zero within the statistical uncertainty, indicating that no statistically significant TRIV signal is observed. The horizontal lines in the figure correspond to $\Delta\chi^2=1.0$ with 68\% confidence level and $\Delta\chi^2=2.71$ with 90\% confidence level for a one-degree-of-freedom profile $\chi^2$. From the intersections with $\Delta\chi^2=2.71$, the following upper limit is obtained as $|W_T|<15\,{\rm eV}$. The value of $E_p$ obtained at the minimum is $E_p=0.792\pm0.006\,{\rm eV}$. The relatively flat behavior of the profile around $W_T \approx 0$ indicates a weak sensitivity to $W_T$ in this region. This can be understood from Eq.~\ref{eq_Imbeta_appro}, where the TRIV contribution enters through the term ${\rm Re}D'{\rm Im}D'$, leading to an approximate quadratic dependence on $W_T$. Figure~\ref{fig:chi2prof}(b) shows the profile $\Delta\chi^2(E_p)$, where two local minima appear around $E_p \approx 0.766\,{\rm eV}$ and $E_p \approx 0.792\,{\rm eV}$. This behavior arises because different parameter combinations yield nearly identical $\chi^2$ values for the shape of the $p$-wave resonance asymmetry. Such parameter degeneracies give rise to a multimodal structure in the $\chi^2$ surface. These two minima do not correspond to distinct resonances, but rather represent local solutions of the fit to the same $^{139}$La $p$-wave resonance. Indeed, no other resonances have been reported in this energy region. The local minimum at $E_p \approx 0.766\,{\rm eV}$ also provides a good fit with $\chi^2/{\rm dof} \approx 0.98$, indicating that it is equally acceptable in describing the data. As shown in the lower panel of Fig.~\ref{fig:chi2prof}(b), previously reported values of $E_p$ exhibit a spread that exceeds their quoted statistical uncertainties. In the present analysis, $E_p$ was treated as a free parameter to conservatively account for this uncertainty. It is important to note that the upper limit on $W_T$ remains stable among the different local solutions of $E_p$, and comparable constraints are obtained in all cases. Therefore, the multimodal structure of $E_p$ and its difference from the literature values do not affect the main conclusion of this study. The limit obtained by minimizing over the full parameter space, $|W_T| < 15\,\rm{eV}$, is slightly weaker than the constraint ($|W_T| < 10\,\rm{eV}$) obtained for the local solution at $E_p\simeq0.766\,{\rm eV}$. This value is therefore adopted as a conservative upper limit that accounts for the uncertainty in $E_p$. Using Eq.~\ref{eq_FSA_La}, the TRIV cross section can be expressed through the optical theorem as $\Delta\sigma_{\not{T}\not{P}}=\frac{4\pi}{k_n}\mathrm{Im}D'$. Substituting the upper limit $|W_T|=15\,{\rm eV}$ and evaluating the resonance-averaged cross section over the energy interval $E_p \pm \Gamma_p$, using the literature value $E_p=0.750\,{\rm eV}$ as a representative resonance energy, yields an upper limit of $|\Delta\sigma_{\not{T}\not{P}}|<8.3\times10^2\,{\rm b}$.

\section{Conclusion}

We have extended the density matrix formalism for polarized-neutron transmission through transversely polarized $^{139}$La by explicitly incorporating the forward scattering amplitude. Applying the formalism to the existing transmission data, we obtained an upper limit of $|W_T| < 15~\mathrm{eV}$ at the 90\% confidence level, corresponding to an order-of-magnitude estimate of $|W_T/W| \lesssim 10^4$ for a representative value of $W \sim 1~\mathrm{meV}$~\cite{Frankle1991,Bowman1990,Fadeev2019}. Although this sensitivity remains many orders of magnitude above that anticipated for future dedicated searches, the present analysis validates the methodology using real experimental data. These results provide practical guidance for optimizing next-generation TRIV experiments. For example, the model study in Ref.~\cite{Bowman2014} projects a statistical sensitivity to $W_T/W$ at the level of $10^{-6}$; see also Ref.~\cite{Flambaum2022} for a complementary statistical treatment and sensitivity scaling.

\section*{Acknowledgment}
This work was financially supported by Japan Society for the Promotion of Science KAKENHI Grants No.~17H02889, No.~19K21047, No.~20K14495, No.~23K13122, and No.~24H00222, and No.~25K01024. C. Auton, J. G. Otero-Munoz, and W. M. Snow acknowledge support from US National Science Foundation (NSF) grant PHY-2209481 and the Indiana University Center for Spacetime Symmetries. J. G. Otero-Munoz also acknowledges the support of the National GEM Consortium and the National Science Foundation AGEP program. C. Auton also acknowledges support from the Japan Society for the Promotion of Science. The authors would like to thank the staﬀ of beamline No.22, RADEN, MLF, and J-PARC for
operating the accelerators and the neutron production target. The neutron experiment was conducted as part of the user program (Proposal No. 2022A0101).
\vspace{0.2cm}

\let\doi\relax

\bibliographystyle{ptephy}
\bibliography{139La}

\appendix

\section{Density matrix formalism for neutron transmission}
\label{appandix_A}

\begin{figure}[h]
	\centering
 	\includegraphics[scale=0.3]{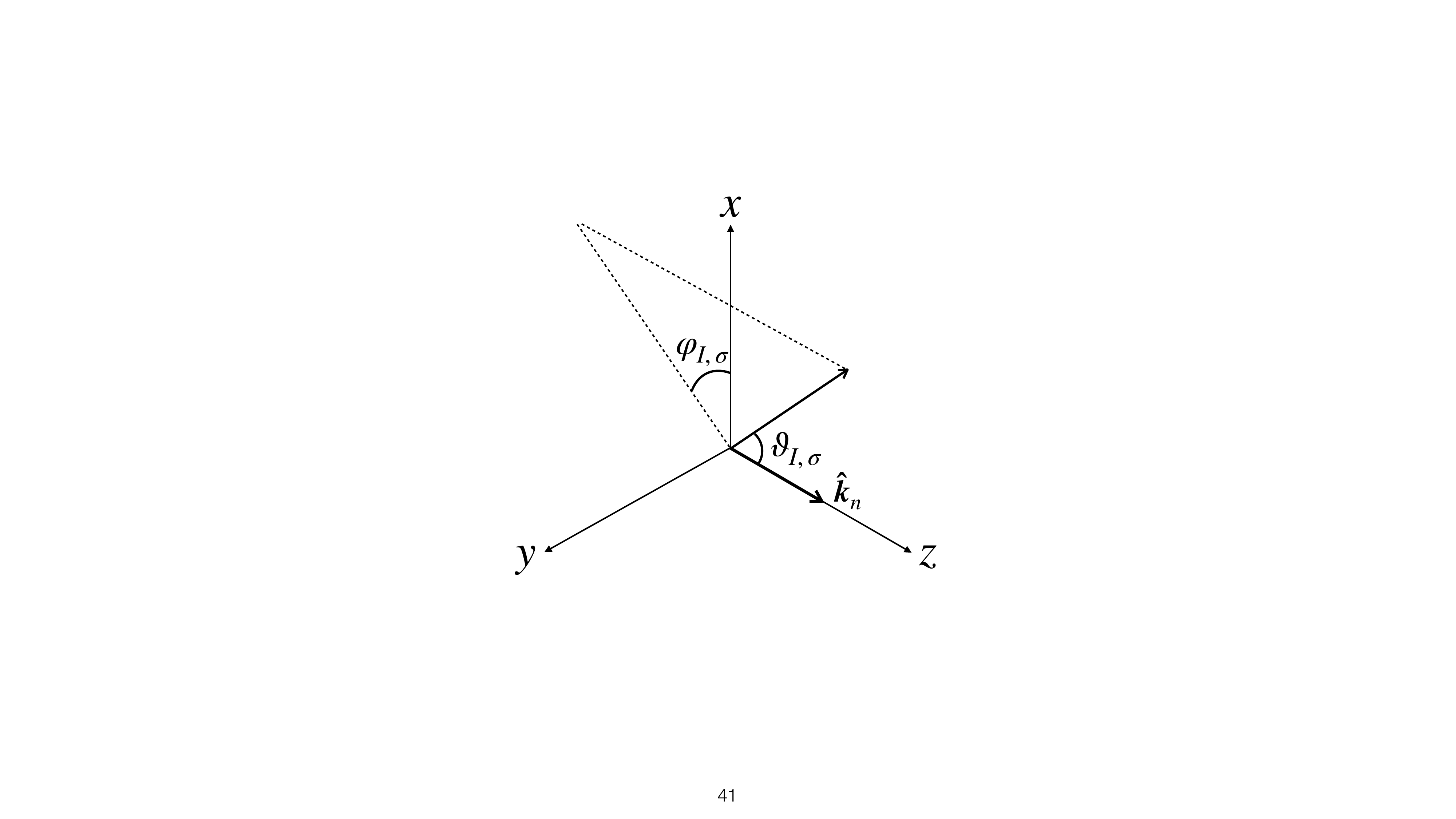}
	\caption{Experimental coordinate system defined by polar angle $\vartheta$ and azimuthal angle $\varphi$. The neutron momentum is fixed along the $z$-axis.}
	\label{fig_expcoord}
\end{figure}

To describe spin-dependent observables in neutron transmission, the density-matrix formalism for neutron spin propagation is employed in the coordinate system shown in Fig.~\ref{fig_expcoord} where the $z$-axis is taken parallel to the momentum of incident neutrons. The orientation of the nuclear spin is specified by the polar and azimuthal angles $(\vartheta_I, \varphi_I)$ with respect to the neutron momentum axis. The corresponding expressions for the neutron wave vector, nuclear spin, and their cross product are given by
\begin{equation}
	\hat{\bm{k}}_n=\begin{pmatrix}0\\0\\1\end{pmatrix},\quad
	\hat{\bm{I}}=\begin{pmatrix}\sin\vartheta_I\cos\varphi_I \\  \sin\vartheta_I\sin\varphi_I \\ \cos\vartheta_I \end{pmatrix},\quad
	\hat{\bm{k}}_n \times \hat{\bm{I}}=\begin{pmatrix} -\sin\vartheta_I\sin\varphi_I \\ \sin\vartheta_I\cos\varphi_I \\0 \end{pmatrix}.
\end{equation}
The effective Hamiltonian in the two-dimensional neutron spin space is expressed as the sum of the Fermi pseudopotential and the interaction with an external magnetic field:
\begin{equation}
	H=-\frac{2\pi\hbar^2}{m_n}\rho f-\bm{\mu}_n\cdot\bm{B}_{\rm ext},
	\label{eq_hamiltonian_n}
\end{equation}
where $\bm{B}_{\rm ext}$ denotes the external magnetic field, which is aligned with the nuclear spin direction. Here, $m_n$ is the neutron mass, $\mu_n$ is the neutron magnetic moment, and $\rho$ is the number density of target nuclei. The non-unitary spin-propagation operator $\mathfrak{S}$ for a neutron propagating through a polarized target of thickness $z$ and velocity $v_n$ is given by~\cite{Lamoreaux1994}
\begin{equation}
	\mathfrak{S} = \exp\left(-\frac{iHz}{v_n\hbar}\right) = \exp\left[iZ\left( \alpha + \bm{\sigma}_n \cdot \tilde{\bm{\beta}} \right) \right],
	\label{eq_evo1}
\end{equation}
where $Z \equiv 2\pi\rho z/k_n$ denotes a wavenumber weighted column density that converts the forward scattering amplitude into dimensionless spin-rotation/spin-dependent attenuation factors in the operator. 
The quantity $\tilde{\bm{\beta}}$ is introduced as the spin-correlation amplitude including an effective scattering length due to the external magnetic field, defined by
\begin{equation}
	\tilde{\bm{\beta}} = \begin{pmatrix}
	(\beta_{\hat{\bm{I}}}-\mu_{\rm eff}B_{\rm ext})\sin\vartheta_I\cos\varphi_I - \beta_{\hat{\bm{k}}_n\times \hat{\bm{I}}}\sin\vartheta_I\sin\varphi_I \\
	(\beta_{\hat{\bm{I}}}-\mu_{\rm eff}B_{\rm ext})\sin\vartheta_I\sin\varphi_I + \beta_{\hat{\bm{k}}_n\times \hat{\bm{I}}}\sin\vartheta_I\cos\varphi_I \\
	\beta_{\hat{\bm{k}}_n}+(\beta_{\hat{\bm{I}}}-\mu_{\rm eff}B_{\rm ext})\cos\vartheta_I
	\end{pmatrix}, \quad \tilde{\beta} = \sqrt{\tilde{\beta}_x^2+\tilde{\beta}_y^2+\tilde{\beta}_z^2},
	\label{eq_multi_beta}
\end{equation}
where $\mu_{\rm eff} \equiv - m_n\mu_n / (2\pi\hbar^2 \rho)$ is the effective magnetic moment in the target.
All three-vectors are expressed as column vectors in the Cartesian basis $(x,y,z)$. Using the exponential identity for spin operators, Eq.~\ref{eq_evo1} becomes
\begin{equation}
	\mathfrak{S} = e^{iZ\alpha} \left[ \cos(Z\tilde{\beta}) + iZ\frac{\sin (Z\tilde{\beta})}{Z\tilde{\beta}} \bm{\sigma}_n\cdot\bm{\tilde{\beta}} \right],
    \label{eq_evo2}
\end{equation}
which yields the matrix form
\begin{equation}
	\mathfrak{S} = \begin{pmatrix}
	A+C & B-iD\\
	B+iD & A-C
	\end{pmatrix},
\end{equation}
with
\begin{equation}
	A = e^{iZ\alpha}\cos(Z\tilde{\beta}),\quad
	B = i Ze^{iZ\alpha} \frac{\sin(Z\tilde{\beta})}{Z\tilde{\beta}} \tilde{\beta}_x,\quad
	C = i Ze^{iZ\alpha} \frac{\sin(Z\tilde{\beta})}{Z\tilde{\beta}} \tilde{\beta}_z,\quad
	D = i Ze^{iZ\alpha} \frac{\sin(Z\tilde{\beta})}{Z\tilde{\beta}} \tilde{\beta}_y.
	\label{eq_ABCD_A}
\end{equation}
The common factor $e^{iZ\alpha}$ governs the overall transmittance. The spin-dependent components ($B$, $C$, $D$) are modulated by the suppression factor $\sin(Z\tilde{\beta})/Z\tilde{\beta}$, representing spin precession about the respective axes. Notably, this suppression is sensitive to the external magnetic field, since the term $\mu_{\rm eff}B_{\rm ext}$ typically dominates over ${\rm Re}\beta_{\hat{\bm{I}}}$.

\section{Expectation value of transmission and observed asymmetry}
\label{Appndix_B}

The neutron transmission probability, normalized to the number of incident neutrons, is obtained from the propagation of the density matrix under the operator $\mathfrak{S}$ for an initially polarized neutron ensemble. Let $(\vartheta_\sigma, \varphi_\sigma)$ denote the polar and azimuthal angles of the initial neutron spin orientation in the coordinate system shown in Fig.~\ref{fig_expcoord}. Then, the initial spin density matrix with polarization $p_n \geq 0$ is given by
\begin{equation}
\hat{\rho}
= \frac{1}{2}
\begin{pmatrix}
1 + p_z & p_x - ip_y \\
p_x + ip_y & 1 - p_z
\end{pmatrix},
\label{eq_density}
\end{equation}
where the polarization components are defined as:
\begin{equation}
p_x = p_n \sin\vartheta_\sigma \cos\varphi_\sigma,\quad
p_y = p_n \sin\vartheta_\sigma \sin\varphi_\sigma,\quad
p_z = p_n \cos\vartheta_\sigma.
\label{eq_density_pxyz}
\end{equation}
Let $\hat{\rho}_+$ and $\hat{\rho}_-$ denote the density matrices for spin-up and spin-down initial states, respectively. The corresponding transmission probabilities $N_+$ and $N_-$ are obtained from Eqs.~\ref{eq_evo2} and \ref{eq_density} as
\begin{equation}
\begin{aligned}
N_+ &= \mathrm{Tr}(\mathfrak{S} \hat{\rho}_+ \mathfrak{S}^\dagger) \\
    &= |A|^2 + |B|^2 + |C|^2 + |D|^2 \\
    &\quad + 2p_x \left[ \mathrm{Re}(A^*B) + \mathrm{Im}(C^*D) \right]
          + 2p_y \left[ \mathrm{Re}(A^*D) + \mathrm{Im}(B^*C) \right]
          + 2p_z \left[ \mathrm{Re}(A^*C) + \mathrm{Im}(D^*B) \right], \\
N_- &= \mathrm{Tr}(\mathfrak{S} \hat{\rho}_- \mathfrak{S}^\dagger) \\
    &= |A|^2 + |B|^2 + |C|^2 + |D|^2 \\
    &\quad - 2p_x \left[ \mathrm{Re}(A^*B) + \mathrm{Im}(C^*D) \right]
          - 2p_y \left[ \mathrm{Re}(A^*D) + \mathrm{Im}(B^*C) \right]
          - 2p_z \left[ \mathrm{Re}(A^*C) + \mathrm{Im}(D^*B) \right].
\end{aligned}
\label{eq_nTrans_Apow}
\end{equation}
The total transmission consists of a spin-independent part (the squared norms) and a spin-dependent part, which reverses sign for opposite spin states. The real parts are associated with direct attenuation from spin-dependent absorption, whereas the imaginary parts arise from secondary polarization effects accumulated during transmission through the polarized target. The observed asymmetry, or analyzing power  $A_I(\vartheta_{kI})$, is defined as
\begin{equation}
\begin{aligned}
A_I(\vartheta_{kI}) &= \frac{N_- - N_+}{N_+ + N_-}\\
&= -\frac{2p_x \left[ \mathrm{Re}(A^*B) + \mathrm{Im}(C^*D) \right]
        + 2p_y \left[ \mathrm{Re}(A^*D) + \mathrm{Im}(B^*C) \right]
        + 2p_z \left[ \mathrm{Re}(A^*C) + \mathrm{Im}(D^*B) \right]}{|A|^2 + |B|^2 + |C|^2 + |D|^2}.
\label{eq_anapow_A}
\end{aligned}
\end{equation}
This expression indicates that, by aligning the neutron spin along an appropriate direction, one can selectively access a desired component of the scattering amplitude in Eq.~\ref{eq_forward}. As pointed out in Ref.~\cite{Bowman2014}, the second-order contributions can be canceled by measuring both the asymmetry and the resulting polarization  $P_I(\vartheta_{kI})$, where the imaginary parts contribute with opposite signs. For instance, choosing the spin orientation $(\vartheta_\sigma, \varphi_\sigma) = (\pi/2, \pi/2)$ to isolate the $\beta_{\hat{\bm{k}}_n\times \hat{\bm{I}}}$ term, one obtains
\begin{equation}
   \frac{A_I(\vartheta_{kI})}{p_{0,a}} + \frac{P_I(\vartheta_{kI})}{p_{0,p}} = -\frac{4\,\mathrm{Re}(A^*D)}{|A|^2 + |B|^2 + |C|^2 + |D|^2},
   \label{eq_AD}
\end{equation}
where $p_{0,a}$ and $p_{0,p}$ represent the analyzing power and polarization efficiencies, respectively. The interference terms $A^*X_i$ and $X_i^*X_j$ for $X_{i(j\ne i)} = B, D, C$ are expressed as
\begin{equation}
\begin{aligned}
	A^*X_i &= \frac{e^{-2\,\mathrm{Im}(Z\alpha)}}{2|\tilde{\beta}|^2} \left[ i\sin(2\,\mathrm{Re}(Z\tilde{\beta})) - \sinh(2\,\mathrm{Im}(Z\tilde{\beta})) \right] \tilde{\beta}^* \tilde{\beta}_i\\
    X^*_iX_j &=\frac{e^{-2\,\mathrm{Im}(Z\alpha)}}{2|\tilde{\beta}|^2}\left[ i\cos(2\,\mathrm{Re}(Z\tilde{\beta}))+ \cosh(2\,\mathrm{Im}(Z\tilde{\beta})) \right] \tilde{\beta}_i^* \tilde{\beta}_j
    \label{eq_AXi_A}
\end{aligned}
\end{equation}
and the sum of squared norms as
\begin{equation}
\begin{aligned}
	&|A|^2 + |B|^2 + |C|^2 + |D|^2\\
	&= \frac{e^{-2\,\mathrm{Im}(Z\alpha)}}{2} \Bigg[
	\cos(2\,\mathrm{Re}(Z\tilde{\beta})) + \cosh(2\,\mathrm{Im}(Z\tilde{\beta}))
	+ \frac{-\cos(2\,\mathrm{Re}(Z\tilde{\beta})) + \cosh(2\,\mathrm{Im}(Z\tilde{\beta}))}{|\tilde{\beta}|^2} \sum_i |\tilde{\beta}_i|^2
	\Bigg],
    \label{eq_ABCD_norm_A}
\end{aligned}
\end{equation}
where $i = x, y, z$ and corresponding $X_i = B, D, C$. In the ideal limit where $|\sin(Z\tilde{\beta})/Z\tilde{\beta}| = 1$ holds (i.e., perfect control over the pseudomagnetic field), Eq.~\ref{eq_AD} reduces to the conventional result:
\begin{equation}
	\frac{A_I(\vartheta_{kI})}{p_{0,a}} + \frac{P_I(\vartheta_{kI})}{p_{0,p}} = -2\tanh(2\,\mathrm{Im}(Z\tilde{\beta}_y)) \approx -4Z \mathrm{Im}\beta_{\hat{\bm{k}}_n\times \hat{\bm{I}}}.
	\label{eq_approAsy}
\end{equation}
This expression demonstrates that the combined asymmetry is directly related to the TRIV observable $\Delta \sigma_{\not{T}\not{P}}$, which constitutes the primary quantity probed by the NOPTREX collaboration.

\end{document}